\documentclass[twocolumn]{aastex62}

\usepackage{amsmath}

\def\lesssim{\mathrel{\hbox{\rlap{\hbox{\lower4pt\hbox{$\sim$}}}\hbox{$<$}}}}
\def\gtrsim{\mathrel{\hbox{\rlap{\hbox{\lower4pt\hbox{$\sim$}}}\hbox{$>$}}}}

\newcommand{\msun}{\,\mbox{M$_{\odot}$}}
\newcommand{\msol}{\,\mbox{M$_{\odot}$}}

\newcommand{\kms}{\mbox{\,$\rm{km}\,s^{-1}$}}

\newcommand{\foii}{\mbox{[O\,{\sc ii}]}}
\newcommand{\foiii}{\mbox{[O\,{\sc iii}]}}
\newcommand{\feii}{\mbox{Fe\,{\sc ii}}}
\newcommand{\Ni}{\mbox{$^{56}$Ni}}

\newcommand{\Cr}{\mbox{$^{48}$Cr}}
\newcommand{\LIIFe}{\mbox{$^{52}$Fe}}

\newcommand{\CaII}{Ca\,{\sc ii}}
\newcommand{\TiII}{Ti\,{\sc ii}}

\newcommand{\SiII}{Si\,{\sc ii}}

\newcommand{\cmg}{\,$\mathrm{cm}^{2}\,\mathrm{g}^{-1}$}
\newcommand{\kmsMpc}{\,$\mathrm{km\,s}^{-1}\,\mathrm{Mpc}^{-1}$}
\newcommand{\ergs}{\,erg\,s$^{\mathrm{-1}}$}

\graphicspath{{./}{figures/}}

\submitjournal{ApJ}

\shorttitle{SN2018kzr: rapidly declining transient}
\shortauthors{McBrien et al.}

\begin{document}

\title{SN2018kzr: a rapidly declining transient from the destruction of a white dwarf}

\correspondingauthor{Owen R. McBrien}
\email{omcbrien02@qub.ac.uk}

\author{Owen R. McBrien}
\affil{Astrophysics Research Centre, School of Mathematics and Physics, Queen’s University Belfast, BT7 1NN, UK}

\author{Stephen J. Smartt}
\affiliation{Astrophysics Research Centre, School of Mathematics and Physics, Queen’s University Belfast, BT7 1NN, UK}

\author{Ting-Wan Chen}
\affiliation{Max-Planck-Insitut f\"ur Extraterrestrische Physik, Giessenbachstra{\ss}e, 85748, Garching bei M\"unchen, Germany}

\author{Cosimo Inserra}
\affiliation{Department of Physics and Astronomy, University of Southampton, Southampton SO17 1BJ, UK}

\author{James H. Gillanders}
\affiliation{Astrophysics Research Centre, School of Mathematics and Physics, Queen’s University Belfast, BT7 1NN, UK}


\author{Stuart A. Sim}
\affiliation{Astrophysics Research Centre, School of Mathematics and Physics, Queen’s University Belfast, BT7 1NN, UK}

\author{Anders Jerkstrand}
\affiliation{Max-Planck-Institut f\"ur Astrophysik, Karl-Schwarzschild-Stra{\ss}e 1, 85748 Garching bei M\"unchen, Germany}

\author{Armin Rest}
\affiliation{Space Telescope Science Institute, 3700 San Martin Drive, Baltimore, MD 21218, USA}

\author{Stefano Valenti}
\affiliation{Department of Physics, University of California, 1 Shields Avenue, Davis, CA 95616-5270, USA}

\author{Rupak Roy}
\affiliation{Inter-University Centre for Astronomy and Astrophysics, Ganeshkhind, Pune 411007, Maharashtra, India}

\author{Mariusz Gromadzki}
\affiliation{Warsaw University Astronomical Observatory, Al. Ujazdowskie 4, 00-478, Warszawa, Poland}

\author{Stefan Taubenberger}
\affiliation{Max-Planck-Institut f\"ur Astrophysik, Karl-Schwarzschild-Stra{\ss}e 1, 85748 Garching bei M\"unchen, Germany}

\author{Andreas Fl\"ors}
\affiliation{Max-Planck-Institut f\"ur Astrophysik, Karl-Schwarzschild-Stra{\ss}e 1, 85748 Garching bei M\"unchen, Germany}
\affiliation{European Southern Observatory, Karl-Schwarzschild-Stra{\ss}e 2, 85748 Garching bei M\"unchen, Germany}

\author{Mark E. Huber}
\affiliation{Institute of Astronomy, University of Hawaii, 2680 Woodlawn Drive, Honolulu, Hawaii 96822, USA}

\author{Ken C. Chambers}
\affiliation{Institute of Astronomy, University of Hawaii, 2680 Woodlawn Drive, Honolulu, Hawaii 96822, USA}

\author{Avishay Gal-Yam}
\affiliation{Department of Particle Physics and Astrophysics, Weizmann Institute of Science, Rehovot 76100, Israel}


\author{David R. Young}
\affiliation{Astrophysics Research Centre, School of Mathematics and Physics, Queen’s University Belfast, BT7 1NN, UK}

\author{Matt Nicholl}
\affiliation{Institute for Astronomy, University of Edinburgh, Royal Observatory, Blackford Hill, EH9 3HJ, UK}
\affiliation{Institute for Gravitational Wave Astronomy, School of Physics and Astronomy, University of Birmingham, Birmingham B15 2TT, UK}

\author{Erkki Kankare}
\affiliation{Tuorla Observatory, Department of Physics and Astronomy, University of Turku, FI-20014 Turku, Finland}

\author{Ken W. Smith}
\affiliation{Astrophysics Research Centre, School of Mathematics and Physics, Queen’s University Belfast, BT7 1NN, UK}

\author{Kate Maguire}
\affiliation{School of Physics, Trinity College Dublin, The University of Dublin, Dublin 2, Ireland}


\author{Ilya Mandel}
\affiliation{Monash Centre for Astrophysics, School of Physics and Astronomy, Monash University, Clayton, Victoria 3800, Australia}
\affiliation{OzGrav, Australian Research Council Centre of Excellence for Gravitational Wave Discovery}
\affiliation{Institute for Gravitational Wave Astronomy, School of Physics and Astronomy, University of Birmingham, Birmingham B15 2TT, UK}

\author{Simon Prentice}
\affiliation{School of Physics, Trinity College Dublin, The University of Dublin, Dublin 2, Ireland}

\author{\'Osmar Rodr\'iguez}
\affiliation{Departamento de Ciencias Fisicas, Universidad Andres Bello, Avda. Republica 252, Santiago, Chile}
\affiliation{Millennium Institute of Astrophysics (MAS), Nuncio Monse\~nor S\'otero Sanz 100, Providencia, Santiago, Chile}

\author{Jonathan Pineda Garcia}
\affiliation{Departamento de Astronom\'ia, Universidad de Chile, Camino El Observatorio 1515, Santiago, Chile}

\author{Claudia P. Guti\'errez}
\affiliation{Department of Physics and Astronomy, University of Southampton, Southampton SO17 1BJ, UK}

\author{Llu\'is Galbany}
\affiliation{Departamento de F\'isica Te\'orica y del Cosmos, Universidad de Granada, E-18071 Granada, Spain}

\author{Cristina Barbarino}
\affiliation{The Oskar Klein Centre, Department of Astronomy, AlbaNova, SE-106 91 Stockholm, Sweden}


\author{Peter S. J. Clark}
\affiliation{Astrophysics Research Centre, School of Mathematics and Physics, Queen’s University Belfast, BT7 1NN, UK}


\author{Jesper Sollerman}
\affiliation{The Oskar Klein Centre, Department of Astronomy, AlbaNova, SE-106 91 Stockholm, Sweden}

\author{Shrinivas R. Kulkarni}
\affiliation{Cahill Centre for Astrophysics, California Institute of Technology, 1200 East California Boulevard, Pasadena, CA 91125, USA}

\author{Kishalay De}
\affiliation{Cahill Centre for Astrophysics, California Institute of Technology, 1200 East California Boulevard, Pasadena, CA 91125, USA}


\author{David A. H. Buckley}    
\affiliation{South African Astronomical Observatory, PO Box 9, Observatory 7935, Cape Town, South Africa}

\author{Arne Rau}   
\affiliation{Max-Planck-Insitut f\"ur Extraterrestrische Physik, Giessenbachstra{\ss}e, 85748, Garching bei M\"unchen, Germany}

\begin{abstract}

We present SN2018kzr, the fastest declining supernova-like transient, second only to the kilonova, AT2017gfo. SN2018kzr is characterized by a peak magnitude of $M_r = -17.98$, peak bolometric luminosity of ${\sim} 1.4 \times 10^{43}$\ergs\ and a rapid decline rate of $0.48 \pm 0.03$\,mag\,d$^{\textrm{-1}}$ in the $r$ band. 
The bolometric luminosity evolves too quickly to be explained by pure \Ni\ heating, necessitating the inclusion of an alternative powering source. Incorporating the spin-down of a magnetized neutron star adequately describes the lightcurve and we
estimate a small ejecta mass of $M_\mathrm{ej} = 0.10 \pm 0.05$\msol.
Our spectral modelling suggests the ejecta is composed of intermediate mass elements including O, Si and Mg and trace amounts of Fe-peak elements, which disfavours a binary neutron star merger.
We discuss three explosion scenarios for SN2018kzr, given the low ejecta mass, intermediate mass element composition and the high likelihood of additional powering - core collapse of an ultra-stripped progenitor, the accretion induced collapse of a white dwarf and the merger of a white dwarf and neutron star. The requirement for an alternative input energy
source favours either the accretion induced collapse with magnetar powering or a white dwarf - neutron star merger with energy from disk wind shocks. 

\end{abstract}

\keywords{supernovae: individual (SN2018kzr) -- stars: white dwarfs -- stars: magnetars}

\section{Introduction} 
\label{sec:introduction}

Within the already diverse range of explosive transients known to exist is a subset of rapidly evolving objects commonly referred to as fast transients. The definition of `fast' has changed over time as more of these objects have been discovered, but in general they display rise and fall times much shorter than for a typical supernova, appearing and fading from view in a matter of weeks. Naturally, to exhibit a more rapid evolution than a typical supernova, a different explosion scenario is needed to explain the event beyond the standard single progenitor scenarios studied. 
Examples of fast transients include Type Iax objects such as SN2002cx \citep{Li2003}, .Ia candidates including SN2002bj and SN2010X \citep{Poznanski2010, Kasliwal2010}, Ca-rich Type I supernovae like SN2005E \citep{Perets2010,Valenti2014} and other fast fading transients interesting in their own right like SN2005ek \citep{Drout2013}, iPTF14gqr \citep{De2018}, iPTF16asu \citep{Whitesides2017} and KSN2015K \citep{Rest2018}.
Events such as these tend to be rarer in occurrence, making up only a small fraction of the normal supernova rate.
The fastest transients yet have been discovered in recent years, with the best examples being those of AT2018cow \citep{Prentice2018} and the more recently discovered SN2019bkc \citep{Chen2019}. The fastest optical transient known is, of course, the kilonova AT2017gfo \citep{LigoVirgo2017,Andreoni2017,Arcavi2017,Coulter2017,Chornock2017,Cowperthwaite2017,Drout2017,Evans2017,Kasliwal2017,Lipunov2017,Nicholl2017,Tanvir2017,Pian2017,Troja2017,Smartt2017,Utsumi2017,Valenti2017}, the result of the radioactive decay of heavy r-process elements synthesised in the merger of two neutron stars \citep{Kasen2017,Metzger2017}.
Modern survey telescopes, with their nightly cadences and wide fields-of-view, are uncovering an increasing number of fast transients like these every year. 


Here we report photometric and spectroscopic observations of SN2018kzr, which was independently discovered by the ATLAS survey \citep{Tonry2018} and the Zwicky Transient Facility \citep[ZTF,][]{Bellm2019}, along with modelling of the bolometric lightcurve and early spectra, and a discussion of plausible explosion scenarios of this object.
Throughout this letter we adopt cosmology of $H_{0} = 70$\kmsMpc, $\Omega_{m} = 0.3$ and $\Omega_{\Lambda} = 0.7$ and assume a foreground reddening of $A_{\mathrm{V}}=0.113$ (NED) alongside the \citet{Cardelli1989} extinction law with $R_{\mathrm{V}}=3.1$. All phases are measured with respect to the ZTF discovery epoch, MJD 58480.422, unless otherwise stated.

\section{Observations} \label{sec:observations}


\subsection{Discovery}

SN2018kzr was independently discovered by both ZTF (as ZTF18adaykvg) and ATLAS (as ATLAS18bchu) within 2 hours of each other on the night of 2018 December 28. ZTF discovered it on MJD 58480.422 at $r=18.58\pm0.11$ \citep{Fremling2018TNS} and it was ingested into the public alerts broker {\sc lasair} \citep{Smith2019}, while ATLAS detected it in a 30 second image on MJD 58480.499 with magnitude $o=18.75\pm0.14$. ATLAS has the closest non-detection in time, with four images taken at a midpoint of MJD 58478.520 ($-$1.902 days) and a combined 3$\sigma$ upper limit of $o > 19.66$.
The rapid rise triggered an ePESSTO \citep{Smartt2015} classification spectrum on MJD 58482.317 (+1.895 days) and again on MJD 58483.247 \citep[+2.825 days,][]{Razza2018ATel,Pineda2018ATel}, which suggested a preliminary Type Ic classification.
It is coincident (0\farcs6 offset) with the blue, $g=20.5$, galaxy SDSS J082853.50+010638.6.

\begin{figure*}
    \centering
    \includegraphics[width=0.8\linewidth]{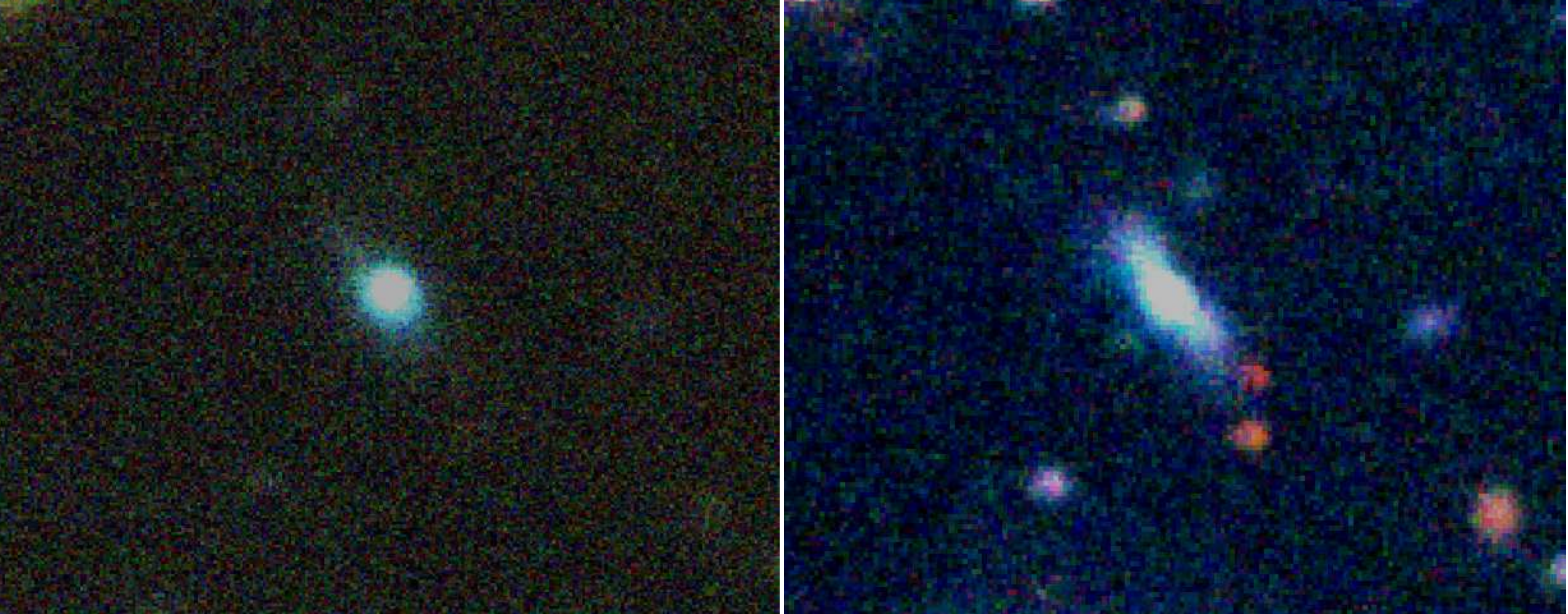}
    \caption{RGB composite images of the host of SN2018kzr, SDSS J082853.50+010638.6. Left: the GROND $gri$ exposures from +3.731 days. Right: the NTT:EFOSC2 $gri$ exposures taken at +68.676 days (Table \ref{tab:photometry}). The host is a blue star-forming galaxy with a bright core.}
    \label{fig:SN2018kzr_host}
\end{figure*}

\subsection{Photometry}

Observations were made over a period of two weeks, during which time the transient faded rapidly (see Table \ref{tab:photometry}). Ground-based $grizJHK$ photometry was collected as part of the GREAT survey \citep{Chen2018} using GROND \citep{Greiner2008}, along with $griz$ photometry from the Liverpool Telescope (LT) and $gri$ photometry from the New Technology Telescope (NTT).
As the transient faded rapidly and was coincident with its host galaxy (Figure \ref{fig:SN2018kzr_host}), difference imaging was essential for all epochs which we carried out using {\sc hotpants} \citep[][]{Becker2015}.
The reference epochs used for the GROND, LT and NTT images are listed in Table \ref{tab:photometry}. Photometry was measured with point-spread-function fitting on the difference images, with the image zero-points set from Pan-STARRS1 reference stars in the field \citep{Chambers2016,Magnier2016phot}.

Eight epochs of UV imaging were taken with Swift. Due to its fast fading, 
it was only recovered in four epochs in $UVW2$, and three in $UVM2$ and $UVW1$. The Swift data are presented in Table \ref{tab:swift_uvot}. These magnitudes have not been host subtracted as host contributions were negligible in the exposures.

\begin{figure}
    \centering
    \includegraphics[width=\linewidth]{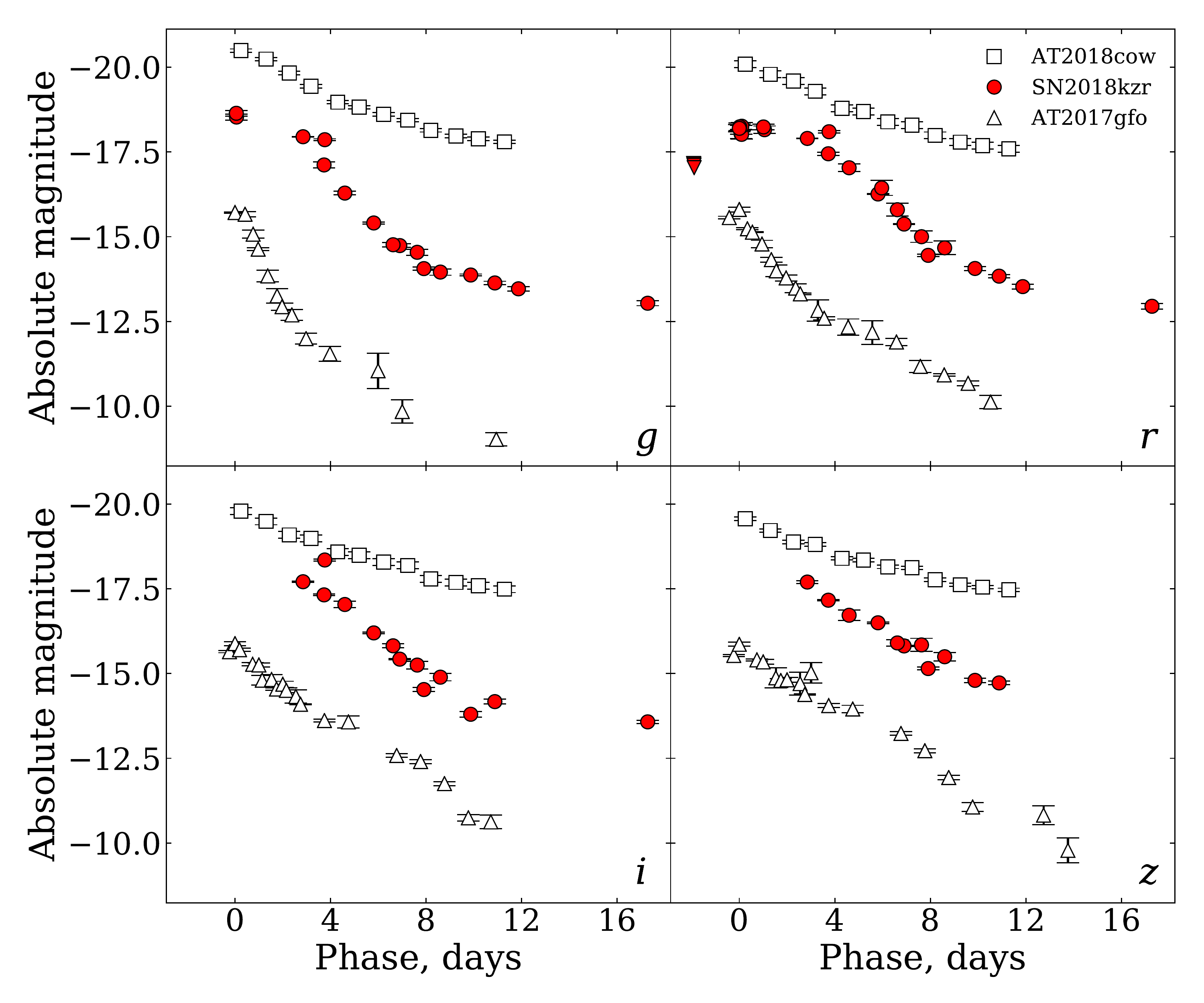}
    \caption{The combined ZTF, ATLAS, GROND, LT and NTT lightcurves compared to the compiled lightcurves of two other notable fast transients - those being AT2018cow \citep{Prentice2018} and the kilonova, AT2017gfo \citep{Andreoni2017,Arcavi2017,Chornock2017,Cowperthwaite2017,Drout2017,Evans2017,Kasliwal2017,Lipunov2017,Nicholl2017,Tanvir2017,Pian2017,Troja2017,Smartt2017,Utsumi2017,Valenti2017}.}
    \label{fig:plotCompareHolistic}
\end{figure}

A strikingly rapid decline was measured across all the $griz$ bands at rates $\Delta g = 0.48 \pm 0.03$\,mag\,d$^{\textrm{-1}}$, $\Delta r = 0.48 \pm 0.03$\,mag\,d$^{\textrm{-1}}$, $\Delta i = 0.54 \pm 0.04$\,mag\,d$^{\textrm{-1}}$, $\Delta z = 0.39 \pm 0.04$\,mag\,d$^{\textrm{-1}}$, all measured over the nine night period for which GROND was observing. 
This is faster than SN2019bkc, it declining at a rate of $\Delta r = 0.41 \pm 0.01$\,mag\,d$^{\textrm{-1}}$ \citep{Chen2019}, which had been the fastest declining supernova-like transient until now. 
The red bands ($i$ and $z$) are similar to the kilonova AT2017gfo (see Figure\,\ref{fig:plotCompareHolistic}). There appears to be no significant near-infrared flux in the GROND $JHK$ images after image subtraction so we do not consider them further here.

\subsection{Spectroscopy}

A total of twelve spectra were taken beginning on MJD 58482.317 (+1.895 days) with the aforementioned initial NTT:EFOSC2 classification spectrum from ePESSTO.
A second NTT:EFOSC2 spectrum was taken on MJD 58483.247 (+2.825 days) along with a third NTT:EFOSC2 spectrum on MJD 58484.172 (+3.750 days) with a broader wavelength coverage ($3400 - 10300$\,\AA\ as opposed to the former $3700 - 9300$\,\AA).
On MJD 58487, three optical to near-infrared spectra were taken by SALT:RSS, VLT:Xshooter and Gemini:GMOS-N, along with a Keck:LRIS spectrum on the subsequent night. A second Gemini:GMOS-N spectrum was taken on MJD 58489.437 (+9.015 days).
Another NTT:EFOSC2 spectrum was obtained on MJD 58490.316 (+9.894 days) but was of a very low signal-to-noise, showing no identifiable emission or absorption features.
One more Keck:LRIS spectrum was taken on MJD 58494.356 (+14.136 days) showing faint emission most notably around 8500\,\AA.
The final spectrum taken was a VLT:Xshooter spectrum from MJD 58525.119 (+44.697 days) which showed narrow nebular emission lines from the host galaxy but no detectable flux from SN2018kzr.
The \foii\ doublet $\lambda \lambda$3726.03, 3926.47 was resolved into two components and a double Gaussian with full width at half maximum $\mathrm{FWHM}=1.7$\,\AA\ was fit to the profiles.
The \foiii\ $\lambda$5006.84 line was also detected and the mean of all three centroids gave $z = 0.05298\pm0.00005$. For the cosmology we adopt, this equates to a luminosity distance of $236$\,Mpc. This VLT:Xshooter spectrum was also used to subtract host continuum flux from the later-time spectra (those from MJD 58487 onward).

\section{Data Analysis} \label{sec:data_analysis}

\subsection{Lightcurve modelling and comparison} \label{subsec:lc_modelling}

\begin{figure*}
    \centering
    \includegraphics[width=0.9\linewidth]{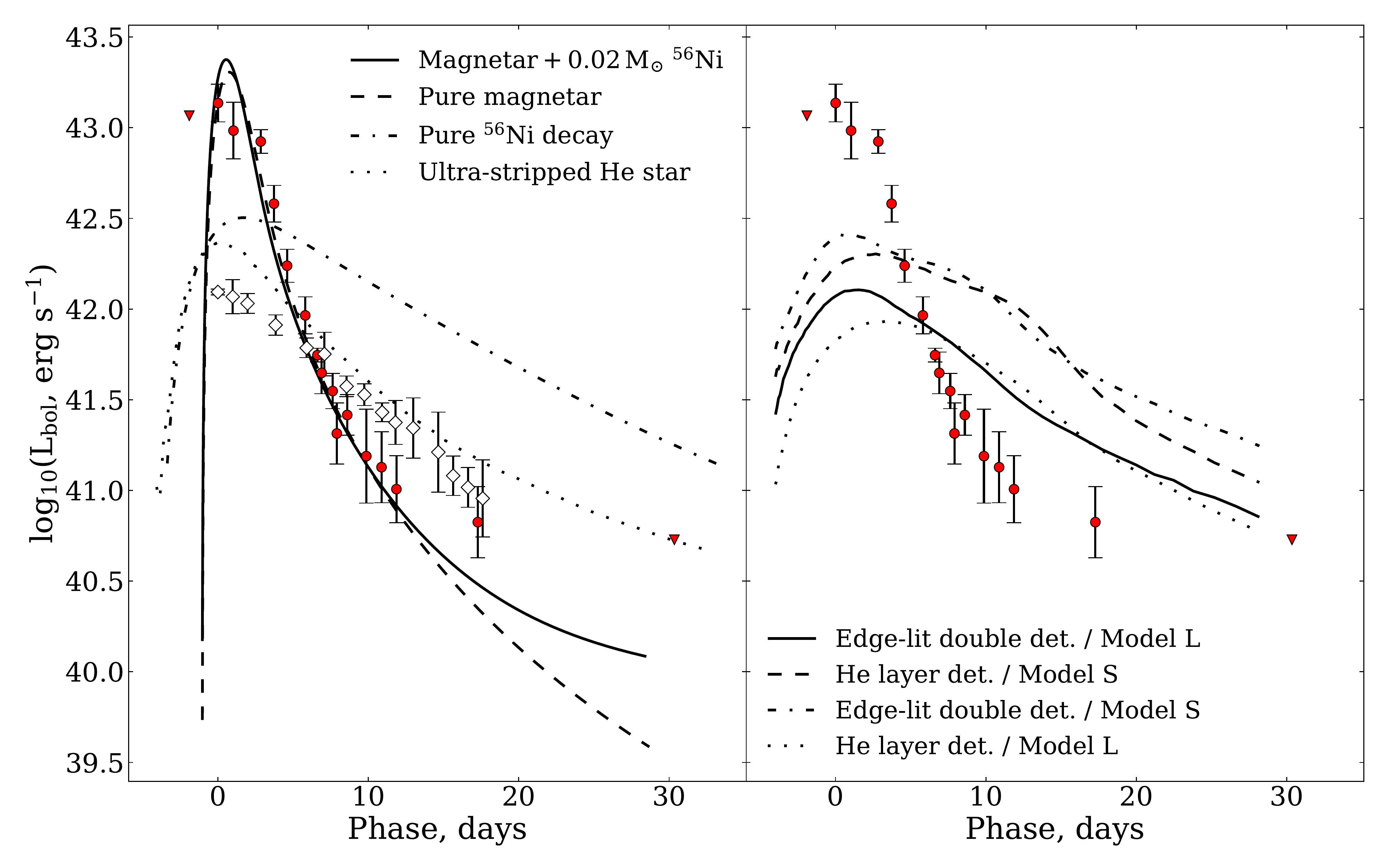}
    \caption{Left: The bolometric lightcurve of SN2018kzr along with several model fits of the powering source including pure \Ni, a mix of \Ni\ heating and magnetar spin-down and the explosion of a stripped He star \citep{Tauris2013ultrastrip}, with the bolometric lightcurve of SN2005ek \citep{Drout2013} for which this model was developed}. Right panel: The bolometric lightcurve of SN2018kzr in comparison to several fast evolving thermonuclear progenitor models detailed in \citet{Sim2012}.
    \label{fig:plotLbolModelCompare}
\end{figure*}

Using the $griz$ photometry (Table \ref{tab:photometry}, Figure \ref{fig:plotCompareHolistic}), a bolometric lightcurve was constructed with \textsc{superol} \citep{Nicholl2018}, which integrates under blackbody fits to the spectral energy distribution estimated at each epoch of observation (Figure \ref{fig:plotLbolModelCompare}).
Based on the Arnett formalism, we may constrain the ejecta mass expected from the opacity, photospheric velocity of the ejecta and an estimate of the rise time of the bolometric lightcurve. Supposing an opacity of $0.1 - 0.2$\cmg, velocity of the order of $10^{4}$\kms\ and a rise time ${<}3$ days, we anticipate an ejecta mass ${\lesssim} 0.1$\msol.

For parameter estimation we have fitted two different powering models, and a combination of both, to the bolometric lightcurve using the formalism and methods described in \cite{Inserra2013}. The powering sources were \Ni\ radioactivity and energy from the spin-down of a magnetic neutron star. In addition we also compare our measured lightcurve to published models of rapidly evolving transients. 
Figure \ref{fig:plotLbolModelCompare} shows the model comparisons, illustrating that the rapid decline rate cannot be fit with a radioactively powered model. To produce a peak luminosity of $L \sim 10^{43}$\,erg\,s$^{-1}$ a mass of 0.17\msol\ is required if $^{56}$Ni is the sole powering source: 

\begin{equation} \label{eq:Lbol_56Ni}
    L_{\rm 56Ni}(t) = 7.8 \times 10^{43}\Bigg(\frac{M_{\rm 56Ni}}{\rm 1M_{\odot}}\Bigg)e^{-t/\tau_{\rm 56Ni}}{\rm ~~~ erg\,s}^{-1}
\end{equation}

Following equation \ref{eq:Lbol_56Ni}, semi-analytical solutions for such a pure \Ni\ model are unable to adequately fit the decline rate as shown in Figure \ref{fig:plotLbolModelCompare}. We show our formal `best fit' model for \Ni\ only powering which has an ejecta mass of 0.28\msol\ assuming an opacity of $\kappa = 0.1$\cmg\ and \Ni\ mass of 0.07\msol. 
Such an ejecta mass has a 5 day rise to peak, a blackbody temperature of $T_{\rm eff} \sim 9000$\,K and would require a velocity of around $20000 - 30000$\kms. This is simply the best fit to the data from a reduced $\chi^2$ statistic. Such a model could be scaled up to fit the peak with a significantly higher mass of \Ni, but declines much too slowly to match the observed data.

The core collapse of an ultra-stripped He star model of \cite{Tauris2013ultrastrip} has been previously applied to rapidly declining transients such as SN2005ek \citep{Drout2013}.
The progenitor transfers material to a compact companion and experiences iron core collapse while only just above the Chandrasekhar limit.  
As can be seen in Figure \ref{fig:plotLbolModelCompare}, even this ultra-stripped model, with $M_{\rm ej}= 0.1$\msol\ and $M_{\rm 56Ni} = 0.05$\msol, does not decline rapidly enough to describe SN2018kzr.
We discuss this explosion scenario in more depth in Section \ref{sec:expl_scenario}.

To further illustrate that rapidly declining models that are \Ni\ powered are inconsistent with the observed data, we show a set of thermonuclear explosion models for low mass carbon-oxygen (CO) white dwarfs (WDs) from \citet{Sim2012} in the right panel of Figure \ref{fig:plotLbolModelCompare}. The \citet{Sim2012} models have a CO core which accretes a sufficiently large helium layer prior to the ignition of core nuclear burning such that the He layer itself instigates a detonation. This primary detonation extends into the CO core wherein a secondary detonation may occur - the Edge-Lit Double Detonation (ELDD) scenario. The primary detonation may, however, be the only detonation to occur, giving the He-layer Detonation (HeD) scenario.
Two sets of models are presented for each scenario, one being the nominated standard system (Model S) with a core mass of $M_{\textrm{CO}} = 0.58$\msun\ and envelope mass of $M_{\textrm{He}} = 0.21$\msun, and another being a specific low mass system (Model L) where the core mass is reduced to $M_{\textrm{CO}} = 0.45$\msun.
The helium shell detonation models (.Ia models) of \citet{Shen2010} are also either too faint, too slowly evolving or too red (see Section 3.2) to be viable explanations.
From this, we disfavour a low mass, thermonuclear explosion, or any type of radioactively powered explosion where the dominant component is \Ni\ as the explosion scenario for SN2018kzr.
We also disfavour powering from other radioactive isotopes, such as \Cr\ or \LIIFe, which may have a shorter lifetime than \Ni\ \citep{Dessart2014}. The energy release per unit mass from the decay of these isotopes is notably lower than that of \Ni\ which would necessitate a larger quantity of each be synthesized compared to the amount of \Ni\ synthesized in order to explain the lightcurve evolution of this object.

An extra powering source is therefore required, and hence we move to testing a model with additional energy from a central engine. We employ a magnetar spin-down component as conceived by \citet{Kasen2010} and \citet{Woosley2010} and further generalised for lightcurve fitting by \citet{Inserra2013}\footnote{ https://bitbucket.org/andersjerkstrand/lcmodels/src/master/}.
Our model supplements \Ni\ decay with powering from the magnetar's rotational kinetic energy as it spins down.
These models assume an explosion energy of $10^{51}$\,erg, a magnetar radiation opacity of 0.01\cmg\ and an electromagnetic radiation opacity of 0.1\cmg. We choose this latter opacity as it is within the limit allowed for electron scattering, assuming it is not influenced by line contributions.
We first considered only the magnetar spin-down component in the absence of \Ni\ powering and found a reasonable fit which implied an ejecta mass of 0.1\msol, along with an initial magnetar spin period of $P=25$\,ms and magnetic field of $B=25 \times 10^{14}$\,G. This fit is shown in Figure \ref{fig:plotLbolModelCompare}. In general, it adequately describes the rapid decline of SN2018kzr but falls below the luminosity of the final data point on the lightcurve. For these fit parameters the spin-down timescale for a magnetar would be approximately 7 days. Hence, given the lifetime of SN2018kzr the input magnetar energy would only decline by a factor of a few. However, the output magnetar energy declines by a factor of nearly 100 in this time, implying the rapid evolution is driven by declining trapping of the magnetar radiation.

It is possible to add a small quantity of \Ni\ to the magnetar model to slow the decline in the tail of the lightcurve and enable a better fit to the late lightcurve.
We observe that 0.02\msol\ of \Ni\ is required to cause a noticeable change in the fit profile, but that this is insufficient for the fit the encompass the final data point.
Further increases in \Ni\ produce less physically plausible fits as the \Ni\ fraction begins to tend to unity.
Furthermore, the bolometric luminosity at this point is uncertain by 0.2 dex. Scaling the +14.136 day Keck:LRIS spectrum to the $gri$ photometry at +16.792 days and integrating the spectral flux gives a luminosity $\mathrm{log}\,L_{\mathrm{bol}} = 40.7$ dex indicating the data may not significantly discrepant from the model.
Both the magnetar only and magnetar supplemented by \Ni\ heating models favour an ejecta temperature in the range of $16000 - 18000$\,K and photospheric velocity of ${\sim} 0.1$\,c at time when the bolometric lightcurve is at peak.

\subsection{Spectral analysis and modelling}

\begin{figure}
    \centering
    \includegraphics[width=\linewidth]{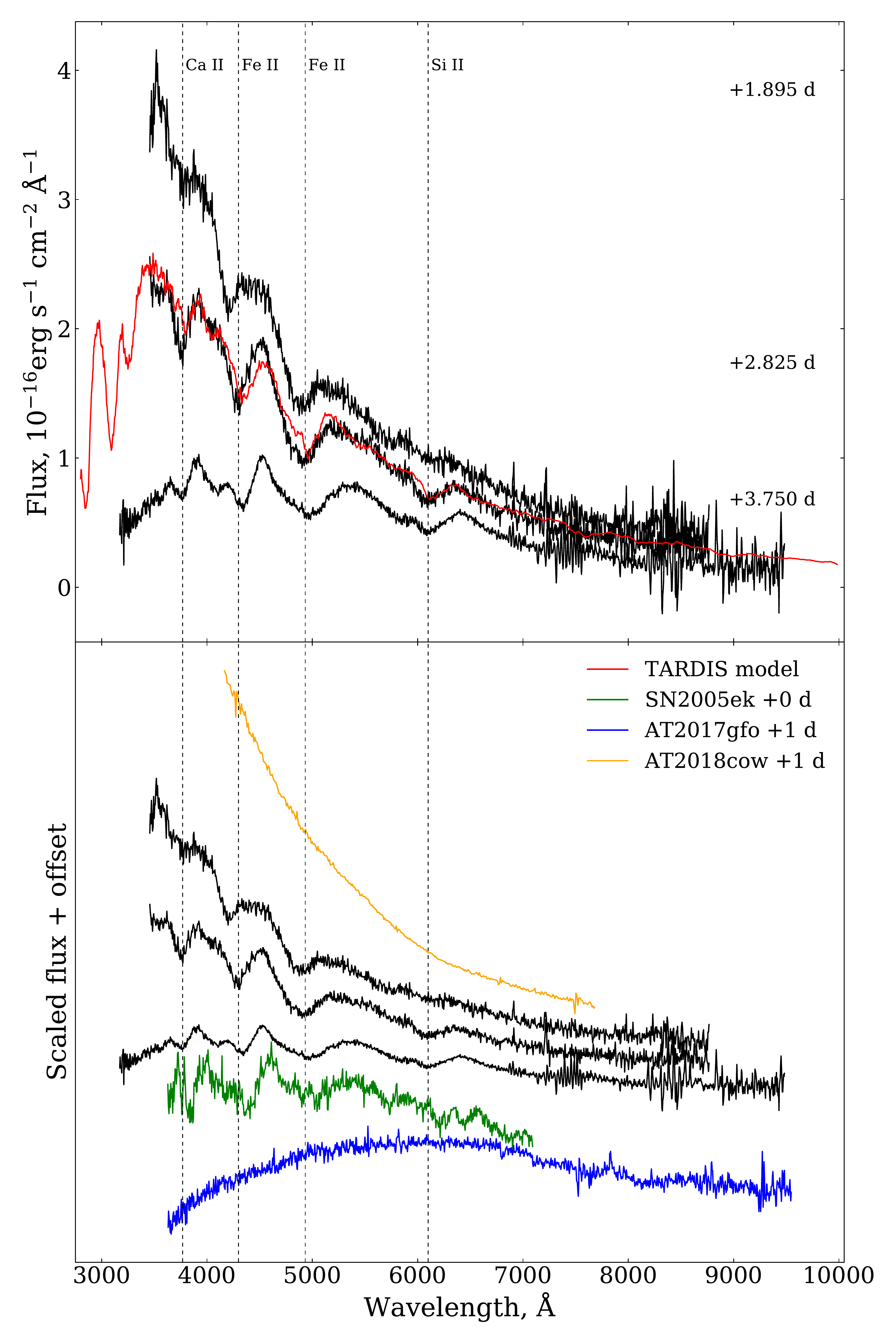}
    \caption{Top: The early spectral sequence of SN2018kzr comprised of the first three NTT spectra with phases shown relative to the ZTF discovery epoch. Overlaid in red on the +2.825 day spectrum is a {\sc TARDIS} \citep{Kerzendorf2014} model investigating the composition of the progenitor. Bottom: The early spectral sequence of SN2018kzr along with several comparison spectra of fast transients including SN2005ek \citep{Drout2013}, AT2017gfo \citep{Smartt2017} and AT2018cow \citep{Prentice2018}. Phases of the comparison spectra are given with respect to the object's maximum light. The spectra have been dereddened and corrected for redshift.}
    \label{fig:plotEarlySpectralSequence}
\end{figure}

Our early spectra were modelled with {\sc TARDIS} \citep{Kerzendorf2014} and a model fit is shown in Figure \ref{fig:plotEarlySpectralSequence} for the NTT:EFOSC2 +2.825 day spectrum. 
There are four strong absorption features with minima at $3900$, $4300$, $5000$, and $6100$\,\AA\, which are reproduced in our model by \CaII, \feii\ and \SiII, with a model velocity of ${\sim}12000$\kms. 

The model is primarily composed of O (${\sim} 75$\%), with significant amounts of intermediate mass elements, primarily Si and Mg (${\sim} 10$\% each), along with some Fe group elements. 
To reproduce the \feii\ features in our observed spectra we require ${\sim} 3$\% of the total ejecta mass to be Fe in our model.
We previously found that 20\% of the ejecta being \Ni\ is required to impact the lightcurve fit, but for this composition if as much of the $5-10\%$ of the ejecta is \Ni\ it begins to present significantly in the spectral model fit. Hence, we disfavour a large amount of \Ni\ in the ejecta.

The temperature, ejecta mass and luminosity required for the spectral fit in Figure \ref{fig:plotEarlySpectralSequence} are consistent with the lightcurve model, with some minor discrepancies.
The model spectrum is 7 days after explosion, whereas the lightcurve fit implies this spectrum should be $4-5$ days after explosion. This may imply the ejecta is not in homologous expansion and given the simplicity of our magnetar model for the lightcurve, where the hydrodynamics of the pulsar wind bubble is not numerically modelled, we do not consider this a serious physical inconsistency.
The ejecta velocity implied by the lightcurve modelling stands at a factor of three greater than that by the spectral modelling. This is likely the result of a longer rise time than is assumed by the lightcurve model, a non-homologous expansion of material or the ejecta being non-spherical.
Further quantitative modelling of all spectra and a more detailed description of the radiative transfer will be presented in a companion paper (Gillanders et al. in prep).

\begin{figure}
    \centering
    \includegraphics[width=\linewidth]{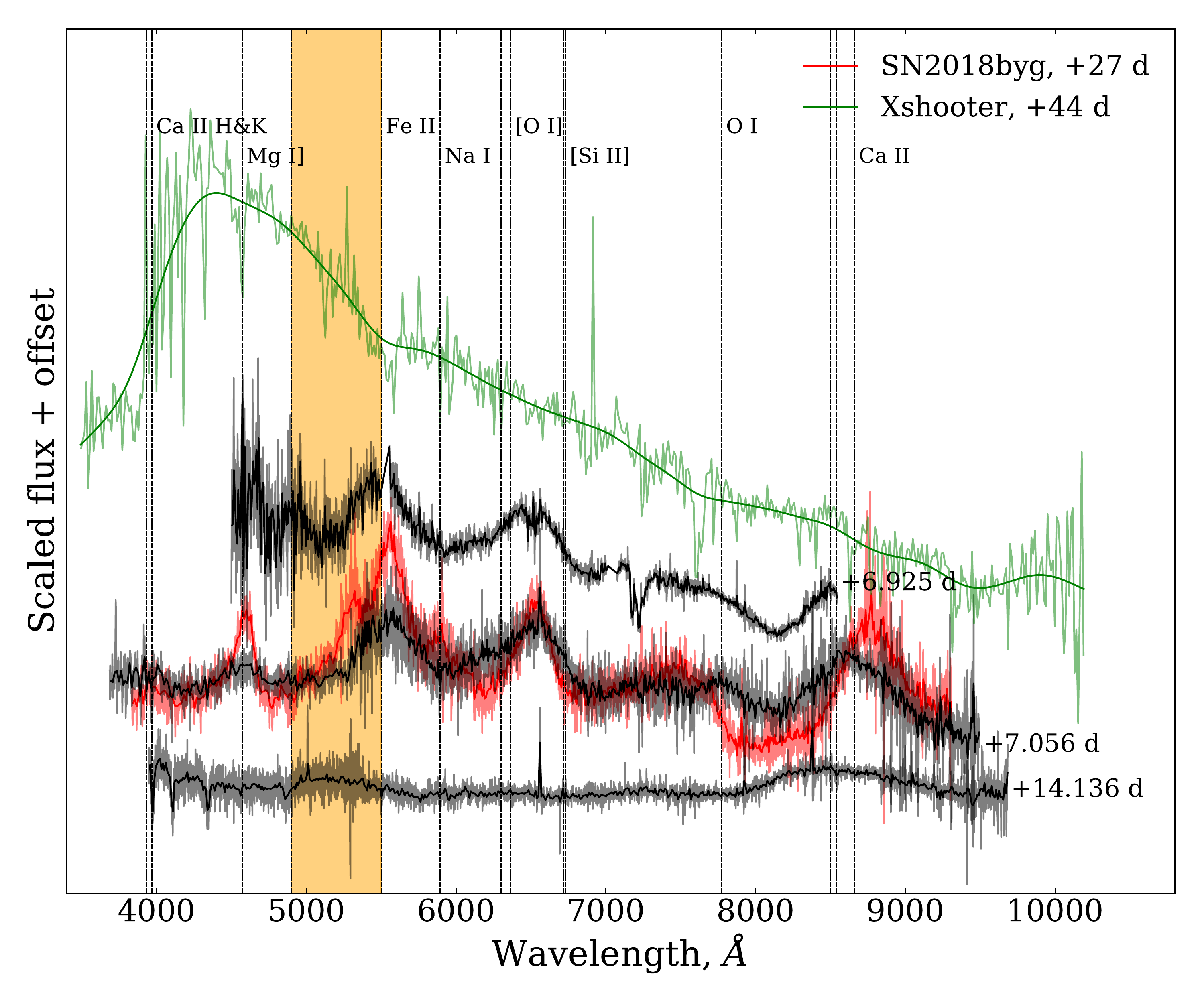}
    \caption{Gemini:GMOS-N spectrum and two Keck:LRIS spectra (black). The final VLT:Xshooter spectrum at +44.697 days has been heavily smoothed (green) and subtracted from these spectra to remove host contamination. The spectra have been rebinned to approximate 5\,\AA\ per pixel resolution. Overlaid on the Keck:LRIS +7.056 day spectrum is a spectrum of SN2018byg at +27 days \citep[red,][]{De2019} from its $r$ band peak highlighting the similarities between these two objects despite their significantly different evolutionary timescales. The shaded region indicates the wavelength range over which \feii\ emission occurs.}
    \label{fig:plotLateSpectralSequence}
\end{figure}

In Figure \ref{fig:plotLateSpectralSequence}, we show the two spectra taken at approximately +7 days from Gemini:GMOS-N and Keck:LRIS (see Table\,\ref{tab:spectra}) as well as the Keck:LRIS spectrum from +14.136 days.
As the transient faded rapidly, contamination from the host galaxy became significant at these epochs. The VLT:Xshooter spectrum from +44.697 days, which is purely host flux, was smoothed and subtracted from the spectra. The spectra were scaled and subtracted such that the final host subtracted spectra matched the difference image photometry through synthetic photometry in the $riz$ bands.
The flux levels are not reliable below $4500$\,\AA\ due to the strong host flux at these wavelengths. 
There is a remarkable similarity between the Keck:LRIS +7.056 day spectrum of SN2018kzr and a SN2018byg spectrum taken +27 days from its $r$ band maximum. SN2018byg, discussed by \citet{De2019}, has been presented as the result of a double detonation of a CO WD surrounded by a He shell, much like the models presented by \citet{Sim2012}. 
The spectra around peak for SN2018byg are noted by \citet{De2019} to show line blanketing from Fe group elements indicative of a large Fe mass in the outermost layers of the ejecta.
The features at $4500$ and $5500$\,\AA\ are weaker in the spectrum of SN2018kzr. \citet{De2018} attribute the features to \CaII\ and \TiII\ in SN2018byg. At this stage, SN2018kzr is entering the nebular phase and further analysis of the ionic species producing these features will be discussed in Gillanders et al. (in prep.).

By +14 days, the $5500$ and $6500$\AA\ features have disappeared, leaving a strong and broad feature centred on $8450$\AA. The obvious candidate is the \CaII\ triplet, however the centroid of the feature is ${\sim}120$\,\AA\ (${\sim} 4200$\kms) offset from its rest wavelength position. 

\section{Explosion mechanism and scenario} \label{sec:expl_scenario}
Our data show SN2018kzr is the fastest declining supernova-like transient apart from the kilonova, AT2017gfo.
We rule out a NS-NS merger for SN2018kzr due to the {\sc TARDIS} spectroscopic model composition which is predominantly intermediate mass elements including O, Mg, Si and Ca, along with a small fraction of Fe.
The lightcurve and spectra cannot be explained only through radioactive powering by \Ni\ and we instead favour a magnetar powering mechanism.
This powering mechanism provides a model which is quantitatively a good fit to the data with an ejecta mass of $M_{\mathrm{ej}} \simeq 0.1$\msol, and a neutron star with period $P \simeq 25$\,ms and magnetic field of $B\simeq25\times10^{14}$\,G.
We disfavour any He-detonation or thermonuclear model due to the fast and luminous light curve, which is physically inconsistent with \Ni\ powering. 
Three possible progenitor scenarios and explosion mechanisms are worth considering that have previously been investigated and predict low mass ejecta with intermediate mass elements and an alternative power source to radioactive decay. These are an ultra-stripped core collapse of a massive star, accretion induced collapse (AIC) of an oxygen-neon (ONe) WD, and a WD-NS merger. 

{\em Ultra-stripped core collapse model:}
The ultra-stripped core collapse model has a He star with total mass before explosion which is only just above the Chandrasekhar limit \citep[e.g. $0.05 - 0.20$\msol,][]{Tauris2015}, due to mass-loss from a common envelope phase and accretion onto a neutron star companion in a tight orbit. The models of \cite{Tauris2013ultrastrip} have successfully reproduced rapidly declining transients such as SN2005ek \citep[][which we show in the left panel of Figure \ref{fig:plotLbolModelCompare} for illustrative purposes]{Drout2013}.
However, such an explosion scenario is unlikely to produce as rapidly rotating a remnant as we present here. For a given He star, the largest component of angular momentum will be held in the envelope rather than the core. Rapid stripping of the envelope via mass transfer to a compact companion does not normally facilitate redistribution of angular momentum to the core.
This is supported by multi-dimensional simulations of \cite{Muller2018} which lead to slowly spinning progenitors, far from the 25\,ms rotation rate required to provide the observed luminosity.

{\em Accretion induced collapse of a white dwarf:}
The accretion induced collapse of an ONe WD has been predicted to lead to a rapidly rotating neutron star in which magnetic fields may be large \citep[up to $10^{15}$\,G,][]{Dessart2007}.
These simulations predict a magnetically enhanced explosion leaving behind a rapidly rotating millisecond pulsar, along with an ejection of ${\sim}0.1$\msol\ of material with only traces of \Ni. The He star + ONe WD binary simulations of \cite{Brooks2017} show that the accretion from a He star companion can lead to an outer layer structure on the ONe WD which is composed of O, Ne, Si and Mg. The $1.0-1.3$\msol\ WD grows, reaching close to the Chandrasekhar limit, which triggers electron capture in the core resulting in AIC.
The composition of the WD calculated by \citet{Brooks2017} is compatible with our estimates from the spectral models. However several simulations have predicted significantly heavier elements should characterise the ejecta of AIC events. \citet{Metzger2009} and \citet{Darbha2010} predict a composition rich in Fe-group elements, while  the \citet{Dessart2007} simulations produce ejecta with a low electron fraction and a composition dominated by elements heavier than Fe. It appears none of these models produce the intermediate mass element composition apparent in our spectra.

{\em White dwarf - neutron star mergers:}
The WD-NS merger scenario involves the production of an accretion disc following the tidal disruption of a sufficiently massive WD as it inspirals with a NS companion.
The disc will be comprised of WD material, provided the WD has mass ${\gtrsim}0.65$\msol \citep{Margalit2016}, and the temperature and mid-plane density are predicted to be high enough to support burning of WD material to higher mass elements \citep{Metzger2012}.
For a CO WD, the ejecta may contain the intermediate mass elements observed in SN2018kzr (O, Si, Mg) in addition to $10^{-3} - 10^{-2}$\msol\ of \Ni\ \citep{Metzger2012}. 
This \Ni\ can only power a faint optical transient of peak luminosity ${\sim}10^{40}$\ergs.
However, high velocity winds from the disk can produce shocks which thermalise the kinetic energy of the winds to power characteristic luminosities of $10^{43}$\ergs \citep{Margalit2016}.
Interestingly the timescale of the powering falls off as $\dot{E} \sim t^{-5/3}$.
This is similar to the magnetar powering function ($t^{-2}$) and hence would likely result in a similar lightcurve. 
\citet{Schwab2016} suggest another possible channel to produce a rapidly rotating neutron star remnant is through the merger of two WDs, potentially avoiding thermonuclear runaway and creating a massive, rapidly rotating WD that will likely collapse \citep{Gvaramadze2019}.

Of these scenarios, we disfavour the ultra-stripped core collapse scenario, owing predominantly to the fact that it would not accommodate such a rapidly rotating neutron star as we are suggesting here. 
We instead favour the AIC or WD-NS merger scenarios as they are consistent with an ejecta mass of $M_\mathrm{ej} = 0.1 \pm 0.05$\msol\ and the requirement from our bolometric lightcurve modelling that the powering mechanism be supplemented by an additional component, likely a rapidly rotating magnetar.. 
Our spectral modelling indicates a composition of primarily intermediate mass elements. In the case of AIC, it is unlikely that this would be observed based on current models \citep{Dessart2007,Metzger2009,Darbha2010}, however such a composition is plausible for a WD-NS merger \citep{Metzger2012}.

\section*{Acknowledgements}
Based in part on observations collected at the European Organisation for Astronomical Research in the Southern Hemisphere, Chile as part of the extended Public ESO Spectroscopic Survey for Transient Objects (ePESSTO), 
program 199.D-0143, the SALT Large Science Programme on transients (2018-2-LSP-001), MNiSW DIR/WK/2016/07, GROND support 
through DFG grant HA 1850/28-1. ATLAS is supported primarily through NASA grant NN12AR55G, 80NSSC18K1575. 
This research made use of \textsc{TARDIS} supported by the Google Summer of Code, ESA's Summer of Code in Space program. 
Funding acknowledgments: STFC  ST/P000312/1 (SJS, SAS);  
Alexander von Humboldt Foundation (TWC);
ERC and H2020 MSC grants [615929, 725161, 758638] (AGY, CPG, KM, LG);
ISF GW excellence center, IMOS, BSF Transformative program, Benoziyo Endowment Fund for the Advancement of Science, Deloro Institute, Veronika A. Rabl Physics Discretionary Fund, Paul and Tina Gardner and the WIS-CIT, Helen and Martin Kimmel Award (AGY); RAS Research Fellowship (MN), Polish NCN MAESTRO grant 2014/14/A/ST9/00121 (MG), IC120009 `Millennium Institute of Astrophysics' of the Iniciativa Cient\'ifica Milenio del Ministerio Econom\'ia, Fomento y Turismo de Chile and CONICYT PAI/INDUSTRIA 79090016 (OR); NRF South Africa (DAHB)
{\sc lasair} is supported by STFC grants ST/N002512/1 and ST/N002520/1.

\facilities{ESO (NTT, VLT), Gemini (GMOS), Keck (LRIS), SALT (RSS), Swift (XRT and UVOT), SALT, GROND, ATLAS, ZTF, LT (IO:O), LCO}




\appendix

\begin{table*}[]
\scriptsize  
\centering
\caption{The $griz$ photometric log of SN2018kzr. All magnitudes, with the exception of ATLAS and ZTF data, were measured following template subtraction of the host galaxy. All phases are presented in the observer frame with respect to the ZTF discovery epoch, MJD 58480.422. ATLAS filter points are converted to $r$ in subsequent plots.
\\$^{a,~b}$ denotes that these are AB magnitudes in the ATLAS $o$ and $c$ filters respectively.
\\$^{c}$ denotes magnitudes obtained via aperture photometry, as opposed to PSF photometry, due to trailing in the input images.}
\begin{tabular}{llcccccr}
\hline
Date                  & MJD            & Phase       & $g$                & $r$                 & $i$                & $z$                & Instrument \\\hline 
20181223 12:23:02 & 58475.516 & $-$4.906 & $>$19.0      & $-$          & $-$          & $-$          & ZTF    \\
20181224 12:11:31 & 58476.508 & $-$3.914 & $-$          & $>$18.90$^{a}$     & $-$          & $-$          & ATLAS  \\
20181226 12:28:48 & 58478.520 & $-$1.902 & $-$          & $>$19.80$^{a}$     & $-$          & $-$          & ATLAS  \\
20181228 10:07:48 & 58480.422 & 0.000    & $-$          & 18.58 $\pm$ 0.11 & $-$          & $-$          & ZTF    \\
20181228 10:34:01 & 58480.440 & 0.018    & $-$          & 18.54 $\pm$ 0.11 & $-$          & $-$          & ZTF    \\
20181228 11:26:59 & 58480.477 & 0.055    & 18.14 $\pm$ 0.08 & $-$          & $-$          & $-$          & ZTF    \\
20181228 11:36:22 & 58480.484 & 0.062    & 18.25 $\pm$ 0.09 & $-$          & $-$          & $-$          & ZTF    \\
20181228 11:58:04 & 58480.499 & 0.077    & $-$          & 18.75 $\pm$ 0.14$^{a}$ & $-$          & $-$          & ATLAS  \\
20181228 12:11:05 & 58480.508 & 0.086    & $-$          & 18.62 $\pm$ 0.13$^{a}$ & $-$          & $-$          & ATLAS  \\
20181228 12:25:10 & 58480.517 & 0.095    & $-$          & 18.76 $\pm$ 0.14$^{a}$ & $-$          & $-$          & ATLAS  \\
20181228 12:38:46 & 58480.527 & 0.105    & $-$          & 18.52 $\pm$ 0.11$^{a}$ & $-$          & $-$          & ATLAS  \\
20181229 10:24:06 & 58481.433 & 1.011    & $-$          & 18.54 $\pm$ 0.08 & $-$          & $-$          & ZTF    \\
20181229 11:28:27 & 58481.478 & 1.056    & $-$          & 18.62 $\pm$ 0.11 & $-$          & $-$          & ZTF    \\
20181230 12:02:52 & 58482.502 & 2.080    & $-$          & 18.68 $\pm$ 0.16$^{a}$ & $-$          & $-$          & ATLAS  \\
20181230 12:15:50 & 58482.511 & 2.089    & $-$          & 18.58 $\pm$ 0.66$^{a}$ & $-$          & $-$          & ATLAS  \\
20181230 12:21:36 & 58482.515 & 2.093    & $-$          & 17.83 $\pm$ 0.55$^{a}$ & $-$          & $-$          & ATLAS  \\
20181230 12:40:19 & 58482.528 & 2.106    & $-$          & 18.70 $\pm$ 0.35$^{a}$ & $-$          & $-$          & ATLAS  \\
20181231 06:34:28 & 58483.274 & 2.852    & 18.83 $\pm$ 0.01$^{c}$ & 18.88 $\pm$ 0.01$^{c}$ & 19.07 $\pm$ 0.02$^{c}$ & 19.08 $\pm$ 0.04$^{c}$ & GROND  \\
20190101 03:40:23 & 58484.153 & 3.731    & 19.66 $\pm$ 0.09 & 19.33 $\pm$ 0.05 & 19.46 $\pm$ 0.02 & 19.62 $\pm$ 0.02 & GROND  \\
20190101 04:19:27 & 58484.180 & 3.758    & 18.92 $\pm$ 0.03 & 18.68 $\pm$ 0.04 & 18.43 $\pm$ 0.03 & $-$             & LCOGT  \\
20190102 00:33:32 & 58485.023 & 4.601	 & 20.49 $\pm$ 0.06 & 19.75 $\pm$ 0.12 & 19.74 $\pm$ 0.10 & 20.06 $\pm$ 0.16 & IO:O    \\
20190103 05:32:53 & 58486.231 & 5.809    & 21.37 $\pm$ 0.03 & 20.52 $\pm$ 0.01 & 20.58 $\pm$ 0.03 & 20.28 $\pm$ 0.02 & GROND  \\
20190103 09:10:04 & 58486.382 & 5.960    & $-$              & 20.34 $\pm$ 0.22 & $-$              & $-$              & P60  \\
20190103 12:04:56 & 58486.503 & 6.081    & $-$          & $>$20.51$^{b}$     & $-$          & $-$          & ATLAS  \\
20190104 00:51:08 & 58487.036 & 6.614	 & 22.02 $\pm$ 0.06	& 20.98 $\pm$ 0.19 & 20.96 $\pm$ 0.06 & 20.88 $\pm$ 0.10 & IO:O \\
20190104 07:42:01 & 58487.321 & 6.899    & 22.04 $\pm$ 0.05 & 21.41 $\pm$ 0.02 & 21.35 $\pm$ 0.02 & 20.96 $\pm$ 0.02 & GROND  \\
20190105 01:04:36 & 58488.045 & 7.623	 & 22.24 $\pm$ 0.09 & 21.78 $\pm$ 0.17 & 21.53 $\pm$ 0.11 & 20.93 $\pm$ 0.20 & IO:O \\
20190105 07:56:24 & 58488.331 & 7.909    & 22.72 $\pm$ 0.05 & 22.33 $\pm$ 0.03 & 22.25 $\pm$ 0.06 & 21.63 $\pm$ 0.04 & GROND  \\
20190106 00:17:30 & 58489.012 & 8.590	 & 22.82 $\pm$ 0.09 & 22.11 $\pm$ 0.20 & 21.87 $\pm$ 0.11 & 21.28 $\pm$ 0.12 & IO:O    \\
20190107 07:09:55 & 58490.299 & 9.877    & 22.91 $\pm$ 0.03 & 22.72 $\pm$ 0.06 & 22.98 $\pm$ 0.08 & 21.98 $\pm$ 0.06 & GROND  \\
20190108 07:21:57 & 58491.307 & 10.885   & 23.14 $\pm$ 0.05 & 22.94 $\pm$ 0.05 & 22.61 $\pm$ 0.07 & 22.06 $\pm$ 0.05 & GROND  \\
20190109 07:10:43 & 58492.299 & 11.877   & 23.32 $\pm$ 0.07 & 23.25 $\pm$ 0.07 & $-$     & $>$21.84     & GROND  \\
20190114 05:08:34 & 58497.214 & 16.792   & 23.74 $\pm$ 0.07 & 23.83 $\pm$ 0.08 & 23.20 $\pm$ 0.05 & $-$          & EFOSC2 \\
20190127 07:11:23 & 58510.300 & 29.878   & $>$24.44         & $>$24.94          & $>$24.11         & $-$          & EFOSC2 \\
20190205 03:08:28 & 58519.131 & 38.709   & ref        & ref        & ref        & ref          & GROND  \\
20190307 02:20:35 & 58549.098 & 68.676   & ref          & ref          & ref          & $-$          & EFOSC2 \\
20190426 20:27:41 & 58599.853 & 119.431 & ref           & ref           & ref           & ref        & IO:O    \\\hline
SDSS DR15   & Host & Model       & 20.58 $\pm$ 0.05 & 20.37 $\pm$ 0.05 & 20.25 $\pm$ 0.08 & 20.35 $\pm$ 0.32 & SDSS   \\
SDSS DR15   & Host & Petrosian   & 20.64 $\pm$ 0.13 & 20.40 $\pm$ 0.09 & 20.17 $\pm$ 0.19  &  - & SDSS   \\
PS1 3$\pi$ & Host  & Kron        & 21.39 $\pm$ 0.08 & 20.64 $\pm$ 0.10 & 20.48 $\pm$ 0.06 & 21.17 $\pm$ 0.29 & PS1  \\
PS1 3$\pi$ & Host  & Aperture    & 21.43 $\pm$ 0.08 & 20.62 $\pm$ 0.08 & 20.59 $\pm$ 0.05 & 21.07 $\pm$ 0.17 & PS1  \\
\hline
\end{tabular}
\label{tab:photometry}
\end{table*}

\begin{table*}[]
\scriptsize  
\centering
\caption{The Swift UVOT photometric log of SN2018kzr. These magnitudes are not host subtracted. All phases are presented in the observer frame with respect to the ZTF discovery epoch, MJD 58480.422.}
\begin{tabular}{llccccccc}
\hline
Date              & MJD       & Phase & $UVW2$             & $UVM2$             & $UVW1$             & $U$                & $B$                & $V$                \\\hline 
20190101 15:54:43 & 58484.663 & 4.241 & 21.55 $\pm$ 0.31 & 21.31 $\pm$ 0.35 & $-$              & $-$              & 19.85 $\pm$ 0.36 & 19.41 $\pm$ 0.49 \\
20190102 05:35:31 & 58485.233 & 4.811 & $-$              & $-$              & $-$              & 21.20 $\pm$ 0.33 & 20.29 $\pm$ 0.35 & $-$              \\
20190102 17:48:28 & 58485.742 & 5.320 & 23.15 $\pm$ 1.74 & $-$              & 21.77 $\pm$ 0.34 & 21.35 $\pm$ 0.45 & 21.34 $\pm$ 1.19 & $-$              \\
20190102 19:24:57 & 58485.809 & 5.387 & $-$              & $-$              & 21.73 $\pm$ 0.34 & $-$              & $-$              & $-$              \\
20190103 15:33:07 & 58486.648 & 6.226 & $-$              & 21.79 $\pm$ 0.26 & $-$              & $-$              & $-$              & $-$              \\
20190103 19:43:40 & 58486.822 & 6.400 & 22.14 $\pm$ 0.19 & 21.83 $\pm$ 0.18 & 21.83 $\pm$ 0.21 & $-$              & $-$              & $-$              \\
20190104 23:36:57 & 58487.984 & 7.562 & $-$              & $-$              & $-$              & $-$              & $-$              & $-$              \\
20190105 22:53:45 & 58488.954 & 8.532 & 21.50 $\pm$ 0.27 & $-$              & $-$              & $-$              & $-$              & $-$              \\  
\hline
\end{tabular}
\label{tab:swift_uvot}
\end{table*}

\begin{table*}[]
\scriptsize  
\centering
\caption{The spectroscopic log of SN2018kzr. All phases are presented in the observer frame with respect to the ZTF discovery epoch, MJD 58480.422.
\\$^{a}$ denotes the resolution is for the Grism\#11 with EFOSC2.
\\$^{b}$ denotes the resolution is for the Grism\#16 with EFOSC2.
\\$^{c}$ denotes the spectra are of low signal, being observed either in poor conditions or with the transient not centred in the slit.}
\begin{tabular}{lllllcc}
\hline
Date              & MJD       & Phase  & Telescope  & Instrument   & Spectral Range     & Spectral Resolution\\\hline
20181230 07:36:05 & 58482.317 & 1.895  & NTT        & EFOSC2       & 3700 - 9300\,\AA   & 355 \\
20181231 05:56:13 & 58483.247 & 2.825  & NTT        & EFOSC2       & 3700 - 9300\,\AA   & 355 \\
20190101 04:06:59 & 58484.172 & 3.750  & NTT        & EFOSC2       & 3400 - 10300\,\AA  & 390$^{a}$, 595$^{b}$ \\
20190104 05:45:10 & 58487.240 & 6.818$^{c}$  & VLT        & Xshooter     & 3100 - 10300\,\AA  & 3300 \\
20190104 08:19:41 & 58487.347 & 6.925  & Gemini     & GMOS-N       & 4200 - 9000\,\AA   & 1918 \\
20190104 11:27:55 & 58487.478 & 7.056  & Keck       & LRIS         & 3000 - 10300\,\AA  & 1050\\
20190104 15:17:01 & 58487.637 & 7.215  & SALT       & RSS          & 3600 - 8300\,\AA   & 1277\\
20190106 10:29:17 & 58489.437 & 9.015$^{c}$  & Gemini     & GMOS-N       & 4200 - 9000\,\AA   & 1918 \\
20190107 07:35:21 & 58490.316 & 9.894$^{c}$  & NTT        & EFOSC2       & 3700 - 9300\,\AA   & 355 \\
20190111 08:33:05 & 58494.356 & 14.136 & Keck       & LRIS         & 3000 - 10300\,\AA  & 1050\\
20190211 02:50:46 & 58525.119 & 44.697 & VLT        & Xshooter     & 3700 - 20700\,\AA  & 3300 \\
\hline
\end{tabular}
\label{tab:spectra}
\end{table*}



\bibliography{ref}

\end{document}